
\input phyzzx

\catcode`\@=11

\def\eqaligntwo#1{\null\,\vcenter{\openup\jot\m@th
\ialign{\strut\hfil
$\displaystyle{##}$&$\displaystyle{{}##}$&$\displaystyle{{}##}$\hfil
\crcr#1\crcr}}\,}
\catcode`\@=12

\def\qg{quantum gravity}
\def\gm{\gamma_{\rm modified}}
\def\L{Liouville}
\def\Tr{{\rm Tr}\,}

\def\mm{matrix model}
\def\NP{{\it Nucl. Phys.\ }}

\def\PL{{\it Phys. Lett.\ }}
\def\PR{{\it Phys. Rev.\ }}
\def\PRL{{\it Phys. Rev. Lett.\ }}

\def\Mod{{\it Mod. Phys. Lett.\ }}

\REF\MAT{V. Kazakov, \PL {\bf 150B} (1985) 282;
J.~Ambj\o rn, B.~Durhuus, and J.~Fr\"ohlich \NP
{\bf B257 }  (1985) 433;
F. David, \NP {\bf B257} (1985) 45; V. Kazakov, I. Kostov
and A. Migdal, \PL {\bf 157B} (1985) 295.}
\REF\sasha{A. M. Polyakov, \PL {\bf 103B} (1981) 207, 211.}
\REF\kpz{V. Knizhnik, A. Polyakov and A. Zamolodchikov,
\Mod {\bf A3} (1988) 819.}
\REF\thorn{T. L. Curtright and C. B. Thorn, \PRL {\bf 48} (1982) 1309. }
\REF\ddk{F. David, \Mod {\bf A3} (1988) 1651;
J. Distler and H. Kawai, \NP {\bf B321} (1989) 509.}
\REF\GM{
D.~J.~Gross and A.~A.~Migdal, \PRL {\bf 64} (1990) 717;
M.~Douglas and S.~Shenker, \NP {\bf B335} (1990) 635;
E.~Brezin and V.~Kazakov, \PL {\bf 236B} (1990) 144.}
\REF\Das{
S.~R.~Das, A.~Dhar, A.~M.~Sengupta, and S.~R.~Wadia,
\Mod {\bf A5} (1990) 1041. }
\REF\abc{ L. Alvarez-Gaum\' e, J.~L. Barbon and C. Crnkovic,
\NP {\bf B394} (1993) 383. }
\REF\korch{ G.~Korchemsky,
\Mod {\bf A7} (1992) 3081; \PL {\bf 296B} (1992)
323. }
\REF\sug{F. Sugino and O. Tsuchiya, UT-Komaba preprint 94-4,
hep-th 9403089.}
\REF\SI{S. Gubser and I. R. Klebanov, Princeton preprint PUPT-1479,
hep-th 9407014.}
\REF\Doug{M. Douglas, \PL {\bf 238B} (1990) 176.}
\REF\dur{B. Durhuus, preprint KUMI-94-2, hep-th 9402052.}
\REF\Joe{J. Polchinski, \NP {\bf B346 } (1990) 253. }
\REF\BK{M. Bershadsky and I. R. Klebanov, \PRL {\bf 65} (1990) 3088.}
\REF\BKN{M. Bershadsky and I. R. Klebanov, \NP {\bf B360} (1991) 559.}
\REF\kdf{
 P. Di Francesco and D. Kutasov, \NP {\bf B342} (1990) 589.}
\REF\ns{N. Seiberg, {\it Prog. Theor. Phys. Suppl.} {\bf 102} (1990) 319;
N. Seiberg and S. Shenker, \PR {\bf D45} (1992) 4581.}
\REF\kkd{V. Kazakov, I. Kostov and J.-M. Daul, \NP {\bf B409} (1993) 311.}
\REF\Wit{E. Witten, \PR {\bf D44} (1991) 314.}
\REF\sw{G. Mandal, A. Sengupta and S. Wadia, \Mod {\bf A6} (1991) 1685;
S. Elizur, A. Forge and E. Rabinovici, \NP {\bf B359} (1991) 581.}
\REF\BHmm{
S.~R. Das, \Mod {\bf A8} (1993) 69;
A. Dhar, G. Mandal and S. Wadia, \Mod {\bf A7} (1992) 3703;
J.~G. Russo, \PL {\bf B300} (1993) 336.}
\REF\antal{A. Jevicki and T. Yoneya,
\NP {\bf B411 } (1994) 64.}

\nopagenumbers
{\baselineskip=16pt
\line{\hfil PUPT-1486}
\line{\hfil July 1994}
\line{\hfil {\tt hep-th/9407167}}
 }

\overfullrule = 0 pt
\normalbaselineskip  = 18pt plus 0.2pt minus 0.1pt
\hsize = 6.2 in
\vsize = 8.9 in
\hoffset =  0.25 in
\title
{TOUCHING RANDOM SURFACES AND LIOUVILLE GRAVITY
}
\author {Igor R. Klebanov
}
\address{\JHL}

\abstract
Large $N$ matrix models modified by terms of the form
$ g(\Tr\Phi^n)^2$ generate random surfaces which touch at isolated
points.
Matrix model results indicate that, as $g$ is increased to a
special value $g_t$, the string susceptibility exponent suddenly
jumps from its conventional value $\gamma$ to ${\gamma\over\gamma-1}$.
We study this effect in \L\ gravity and attribute it to
a change of the interaction term from
$O e^{\alpha_+ \phi}$ for $g<g_t$ to $O e^{\alpha_- \phi}$
for $g=g_t$ ($\alpha_+$ and $\alpha_-$ are the two roots
of the conformal invariance condition for the \L\ dressing of
a matter operator $O$). Thus, the new critical behavior
is explained by the unconventional branch
of \L\ dressing in the action.

\endpage

\pagenumbers
\chapter{Introduction}

A remarkable aspect of the recent developments in
two-dimensional \qg\ has been an interplay between discretized [\MAT]
and continuum [\sasha-\ddk] approaches to the problem. The discretized
approach, implemented mainly via elegant matrix model
techniques, has so far proven more powerful [\GM]. However, the
full significance of the matrix model results usually becomes
apparent only after they are translated into the continuum language
of Liouville gravity. By now there exists a considerable amount of evidence
that the discretized and continuum approaches are indeed equivalent,
but a general demonstration of this has not been found.
We believe that much more can be learned from comparing
the two approaches to two-dimensional \qg.

While the \mm s which generate conventional discretized random surfaces
have been investigated quite thoroughly, there exists a new class
of \mm s where only some partial results are available. These models
describe random surfaces which are allowed to touch each other
at isolated points. This is implemented by adding a term of the form
$ g(\Tr\Phi^n)^2$ to the action of an $N\times N$ hermitian matrix $\Phi$.
\foot{$n$ is often set equal to 2, but all finite $n$ are expected to
lead to the same universal behavior.}
The first matrix integral of this kind,
$$ Z=\int {\cal D} \Phi \, e^{-N \,
   \left[ \Tr \left( \Phi^2 -
     \lambda \Phi^4 \right) -
    {g \over 2 N} \left( \Tr \Phi^2 \right)^2 \right]} \ ,    \eqn\eq
$$
was introduced in ref. [\Das]. The free energy,
$$ F={\log Z\over N^2}\ ,$$
can be expanded in powers of $1/N^2$,
$$ F= F_0 + F_1 N^{-2}+ F_2 N^{-4}+\ldots $$
Each term in this expansion has an interesting geometrical
interpretation.
Feynman graphs of the perturbation theory in $\lambda$
generate the usual connected closed random surfaces, while the
$g (\Tr \Phi^2)^2$ term can glue a pair of such surfaces together at a
point.  This point can be resolved into a tiny neck (a wormhole), so that
the network of such touching surfaces can be assigned an overall genus.
Thus, $F_0$ gives the sum over all such surfaces of overall genus zero
(they look like trees of spherical bubbles such that any two bubbles touch at
most once, and a bubble is not allowed to touch itself).
In general, $F_G$ is the sum over all surfaces of overall genus $G$.

The authors of ref. [\Das] found a critical line in the $(\lambda, g)$
plane where the free energy becomes singular.
For a fixed $g$, $F_0 (\lambda)$ becomes singular at
$\lambda=\lambda_c (g)$ on the critical line. There exists a
critical value $g_t$ such that, for $g<g_t$, the singularity
is characterized by $\gamma=-1/2$, \ie
$$ F_0 (\lambda) \sim (\lambda_c -\lambda)^{2-\gamma}
\sim (\lambda_c -\lambda)^{5/2 } \ .
$$
In this phase
the touching of random surfaces is irrelevant and one finds the
conventional $c=0$ behavior.  For $g>g_t$, on the other hand,
$\gamma=1/2$, and one finds branched polymer behavior, which is
dominated by the touching.  Most interestingly, for $g=g_t$, the authors of
ref. [\Das] found a new type of critical behavior with $\gamma=1/3$.
This is the first example of a \mm\ where new critical behavior
occurs due to fine-tuned touching interactions. We will generally
refer to such new critical points as the modified \mm s.

The results above have been generalized to the $k$-th multicritical
one-matrix model [\abc,\korch],
$$ Z_k=\int {\cal D} \Phi \, e^{-N \,
   \left[ \Tr V_k(\Phi) -
    {g \over N} \left( \Tr \Phi^2 \right)^2 \right]} \ ,    \eqn\eq
$$
where
$$ V_k(\Phi)= \sum_{i=1}^k t_i\Phi^{2i}
\ .$$
For $g=0$, the parameters of the potential can be fine-tuned
to give scaling behavior with $\gamma=-1/k$. As we increase $g$, then
for some $g=g_t$ the scaling exponent suddenly jumps [\abc,\korch] to
$\gm=1/(k+1)$. These values of $\gamma$ are puzzling because they are positive;
in \mm s without the touching interactions only $\gamma\leq 0$ have been
found.

Another piece of the puzzle is provided by the modified $c=1$ \mm,
where a fine-tuning of $g$ also leads to new critical behavior [\sug,\SI],
$$ F_0\sim \Delta^2 \log\Delta\ ,\qquad\qquad \Delta=\lambda_c-\lambda
\ .\eqn\new$$
This should be contrasted with the conventional $c=1$
scaling, $F_0\sim \Delta^2/ \log\Delta$. The sum over spherical
surfaces of fixed area $A$, obtained by an inverse Laplace transform
of eq. \new, scales as $1/A^3$.
In other words, the modified $c=1$ \mm\ has no scaling violations.
An explanation of this effect in terms of \L\ gravity was proposed
in ref. [\SI]. There it was argued that, while the conventional
$c=1$ scaling corresponds to the \L\ potential $\sim\phi e^{-\sqrt 2\phi}$,
at the new critical point the potential is $\sim e^{-\sqrt 2\phi}$
instead.

In this paper we present a \L\ gravity explanation of all the
new critical exponents obtained in the modified \mm s. This explanation
is surprisingly simple and amounts to picking the unconventional branch
in the gravitational dressing of the \L\ potential.
For all the conventional \mm s describing $(p, q)$ minimal
models coupled to gravity, the correct scaling follows from the
\L\ interaction of the form
$$\eqalign{&\Delta \int d^2\sigma O_{\min} e^{\alpha_+\phi}\ ,\cr
&\alpha_+=
{1\over 2\sqrt 3}(\sqrt {1-c+24h_{\min}}-
\sqrt{25-c})
\cr}
$$
where $O_{\min}$ is the matter primary field of the lowest dimension,
$$h_{\min}={1-(p-q)^2\over 4pq}\ .\eqn\ldim$$
 We will argue that the effect of fine-tuning the
touching interaction is to replace the \L\ potential by
$$\eqalign{&\Delta \int d^2\sigma O_{\min} e^{\alpha_-\phi}\ ,\cr
&\alpha_-=
-{1\over 2\sqrt 3}(\sqrt {1-c+24h_{\min}}+
\sqrt{25-c}) \ .
\cr}
$$
It follows that the modified \mm s do not correspond to $c>1$ string
theories; they are simply new solutions of $c\leq 1$ string
theories!

The structure of the paper is as follows. In section 2 we
present the details of our \L\ gravity arguments.
We reproduce the known \mm\ results and make predictions for new calculations.
In section 3 we confirm some of these predictions by finding
the scaling behavior in the modified $(p, p+1)$ \mm s. In section 4 we
discuss the directions for future work.

\chapter{Fine-tuning in Liouville gravity}

Let us consider $(p, q)$ conformal minimal models coupled to
\qg. In the conformal gauge, the sum over surfaces of genus $G$
is given by the path integral [\ddk]
$$F_G=\int d\tau\int [d\Psi][d\phi][db][dc]
e^{-S_\Psi-S_\phi-S_{b,c}} \ ,\eqn\minpath$$
where $S_\Psi$ is the matter action, $S_{b, c}$ is the standard ghost
action, and $\tau$ collectively denotes the moduli.
 The action of the Liouville field is
$$S_\phi
={1\over 8\pi}\int d^2\sigma
\biggl(\partial_a \phi\partial^a\phi- Q\hat R \phi+ O(\Psi) f(\phi)
\biggr)\ ,
\eqn\lact$$
where
$$ Q=\sqrt{25-c\over 3} $$
and $O(\Psi)$ is a matter primary field of dimension $h$. Without extra
fine-tuning, the \L\ potential couples to the lowest dimension primary
$O_{\min}(\Psi)$. It is not hard to work with a general
primary field $O$, although occasionally we will specify our formulae to
$O_{\min}$. The gravitational dressing function $f(\phi)$ is expected
to have the large $\phi$ asymptotic form
$$ f(\phi)\to A e^{\alpha_+ \phi}+
B e^{\alpha_- \phi}
\eqn\asymp$$
where
$$\alpha_\pm=-{Q\over 2}\pm \sqrt {{Q^2\over 4}-2+2h} \eqn\solut$$
are the two solutions of the equation
$$ 2h-\alpha(\alpha+Q) =2 \ .
$$
This equation
guarantees the conformal invariance of \L\ theory in the weakly
interacting region of large $\phi$.
If $A>0$ then, as $\phi$ increases, the first term in eq.
\asymp\ rapidly becomes dominant. Furthermore, large values of $\phi$ are
important in the path integral because they are not suppressed by the
\L\ potential. Therefore, in the generic case $A>0$, we may approximate
$$ f(\phi)\sim \Delta
e^{\alpha_+ \phi}\ .
$$
Applying now the analysis of
ref. [\ddk] we find that the sum over surfaces of genus $G$
obeys the scaling law
$${\partial^2 F_G\over \partial\Delta^2} \sim {1\over \Delta^{2G+\gamma (1-G)}}
$$
where the string susceptibility exponent is
$$\gamma= 2+ {Q\over\alpha_+}\ .\eqn\geng$$

Let us specify now to the case $O= O_{\min}$.
Using eq. \ldim\ and the formula for the central charge,
$$ c=1- 6{(p-q)^2\over pq} \ ,$$
we obtain
$$\eqalign{& Q=\sqrt 2 {p+q\over\sqrt{pq}}\ ,\cr
& \alpha_+= -
{p+q-1\over\sqrt{2pq}}\ .\cr
}\eqn\aplus $$
Therefore,
the string susceptibility exponent is
$$\gamma= -{2\over p+q-1}\eqn\nft$$
which agrees with the conventional \mm\ results [\Doug].

Our main observation is that, by fine-tuning the theory,
we should be able to reach a phase where the dressing function is
given by eq. \asymp\ with $A=0$, \ie\
$$ f(\phi)\sim \Delta
e^{\alpha_- \phi}
\ .\eqn\moddress$$
We believe that, in the language of the \mm s, this fine-tuning
is achieved by setting $g$, the coupling constant for touching interactions,
to $g_t$. Indeed, the touching interactions add tiny wormholes to
surface geometry and, therefore, modify the ultraviolet (large $\phi$)
structure of the theory. It is reasonable that, by fine-tuning the ultraviolet
boundary conditions, we may find a solution for the gravitational dressing
with $A=0$. While the precise mechanism for this is not entirely clear,
we will simply check that the new critical behavior in the modified
\mm s corresponds to the unconventional gravitational dressing, eq.
\moddress.

The calculation of string susceptibility proceeds analogously to ref.
[\ddk], and we find a modified string susceptibility exponent
$$\gm= 2+ {Q\over\alpha_-}\ .\eqn\newg$$
Using eqs. \solut\ and \geng, it is easy to establish that $\gm$ and $\gamma$
are related by
$$ {1\over \gm -2} +{1\over \gamma -2}=-1\ ,
$$
which implies
$$\gm ={\gamma\over\gamma-1}
\ .\eqn\there
$$
This is a completely general relation, independent of which
operator $O$ enters the \L\ potential.
One easily sees that, if $\gamma$ is negative, then $\gm$ is positive.
Thus, positive $\gm$ arise naturally in \L\ gravity! It is interesting that
eq. \there\ was recently obtained by Durhuus [\dur] on the basis of certain
assumptions about random surfaces coupled to spin systems.
Using purely combinatorial arguments
he argued that, given a theory with scaling exponent $\gamma$,
one should be able to construct a theory with scaling exponent
${\gamma\over\gamma-1}$. We have shown how the modified
scaling behavior, eq. \there, arises in \L\ gravity.
We have also proposed a connection between the fine-tuning
in modified \mm s and in \L\ gravity.

Specifying the theory to the case $O= O_{\min}$, we find
$$
\alpha_-= -
{p+q+1\over\sqrt{2pq}}\ .
\eqn\aminus $$
Eq. \newg\ now gives
$$\gm= {2\over p+q+1}\ .\eqn\ft$$
Let us compare this with the known results from the
modified \mm s. Consider, for instance, the $k$-th multicritical
one-matrix model which corresponds to the $(2, 2k-1)$ minimal model
coupled to gravity.
Without the touching interactions, $\gamma=-1/k$ in agreement with eq.
\nft. Eq. \ft\ predicts that, after a fine-tuning of the touching
interactions, $\gamma$ should jump to $1/(k+1)$. This is precisely what
happens [\Das,\abc,\korch]. In our opinion, this provides serious evidence
in favor of our interpretation of \L\ theory.

A similar argument applies to the $c=1$ model, which is special
because $\alpha_+=\alpha_-=\sqrt 2$. As a result, the
\L\ interaction has the form [\Joe]
$$\eqalign{ &\int d^2\sigma T(\phi)\ ,\cr
& T(\phi)\to A\phi e^{-\sqrt 2\phi}
+B e^{-\sqrt 2\phi} \ .\cr }
$$
Without fine-tuning, $A>0$, and the first term dominates for
large $\phi$ giving rise to scaling violations [\Joe].
If, however, we reach a phase with $A=0$, then the  usual DDK
analysis applies, and we find $\gamma=0$ with no scaling
violations [\SI]. This is precisely the new scaling behavior found
in the $c=1$ \mm\ modified by the touching interactions [\sug,\SI].

The \L\ approach is valuable not only in providing a string theoretic framework
for the \mm\ results.
Some calculations can be performed very efficiently starting directly
from the \L\ path integral. Perhaps the simplest such calculations is
the sum over surfaces of genus 1. In ref. [\BK,\BKN] this path integral
was calculated exactly, with the result
$$F_1=
{(p-1)(q-1)\over 24\sqrt{2pq} |\alpha|} |\log\Delta|\ .\eqn\precise $$
For the conventional $(p, q)$ models coupled to gravity we use
$\alpha=\alpha_+$, eq. \aplus, and arrive at
$$F_1=
{(p-1)(q-1)\over 24(p+q-1)} |\log\Delta|\ ,\eqn\eq $$
which agrees with various \mm\ results [\kdf].\foot{Results from \mm s
with even potentials need to be divided by 2 to eliminate
overcounting.}

For the $(p, q)$ models fine-tuned in the sense described above
we may try to use
$\alpha=\alpha_-$, eq. \aminus, in eq. \precise. Then we find
a contribution to the genus 1 free energy,
$$F_1^{\rm modified}=
{(p-1)(q-1)\over 24(p+q+1)} |\log\Delta|\ .\eqn\modtor $$
In ref. [\abc] the scaling of
$F_1^{\rm modified}$ was studied in multicritical one-matrix models
and was found to be $\sim |\log\Delta|$, in agreement with eq. \modtor.
However, the dependence of the normalization on $p$ and $q$ has not
been calculated in the \mm s. Unfortunately, eq. \modtor\ is not
the complete prediction of \L\ theory.
Since the $\alpha_-$ dressed operators are ``macroscopic'',
with wave functions peaked in the strong coupling region [\ns],
they are sensitive to non-trivial string loop corrections.
Such corrections, which probably alter the coefficient of $|\log\Delta|$ in
eq. \modtor, require a separate investigation.

\chapter{Modified two-matrix models}

In the previous section we proposed a \L\ gravity formulation of
the \mm s with fine-tuned touching interactions and obtained a number
of non-trivial predictions. In this section we study such modified
\mm s for the $(p, p+1)$ conformal minimal models coupled to gravity.
We calculate the sum over genus zero surfaces, $F_0 (\Delta)$, and,
after the fine-tuning, find $\gm = {1\over p+1}$, in agreement with
eq. \ft.

As a warm-up, we repeat the calculation for pure gravity ($p=2$),
which may be described by the one-matrix model
$$ Z=\int {\cal D} \Phi \, e^{-N \,
   \left[ \Tr \left( \Phi^2 -
     \lambda \Phi^4 \right) -
    {g \over 2 N} \left( \Tr \Phi^4 \right)^2 \right]} \ .    \eqn\om
$$
Comparing with ref. [\Das], we have replaced $\left( \Tr \Phi^2 \right)^2$
by $\left( \Tr \Phi^4 \right)^2$. This makes the calculation a bit simpler
but, as expected, results in the same universal behavior. We will
use a self-consistent method,
analogous to Hartree-Fock calculations, to analyze eq. \om.
For the purpose of finding the free energy to
the leading order in $N$, it is permissible to make a replacement [\SI]
$$\eqalign{&\left( \Tr \Phi^4 \right)^2 \to
   2 N c\Tr \Phi^4 - Nc^2 \ , \cr
 & c = \left\langle {1 \over N} \Tr \Phi^4 \right\rangle \ . \cr
}
   \eqn\Replace
$$
Substituting this into \om, we arrive at an auxiliary one-matrix model
with coupling constant $\kappa=\lambda+ gc$ and no touching
interactions. Thus, the self-consistency condition has the form
$$
 c = \left\langle {1 \over N} \Tr \Phi^4 \right\rangle \ (\kappa)
\ .\eqn\sc$$
The right-hand side is simply the puncture one-point function in
the conventional one-matrix model with a $\kappa \Tr \Phi^4$ interaction.
Using well-known results, we find that eq. \sc\ becomes
$$ c=a_1-a_2 (\kappa_c-\kappa)+ a_3 (\kappa_c-\kappa)^{3/2}+\ldots
$$
where $a_i$ and $\kappa_c$
are positive constants. Differentiating this with respect to
$\lambda$, we find
$$ {\partial c\over \partial\lambda} (1-a_2 g+{3\over 2} g
a_3 (\kappa_c-\kappa)^{1/2})=a_2- {3\over 2} a_3 (\kappa_c-\kappa)^{1/2}
$$
(from here on we retain only the leading singular terms).
For $g<1/a_2$, ${\partial c\over \partial\lambda}$ is finite at
the critical point $\kappa=\kappa_c$, and the scaling with $\gamma=-1/2$
follows. If, however, we fine-tune $g=1/a_2$, then
$$ {\partial c\over \partial\lambda}= {2\over 3}
{a_2^2\over a_3 (\kappa_c-\kappa)^{1/2}}=
{2\over 3}
{a_2^{5/2}\over a_3 (a_1-c)^{1/2}}
\ .$$
Setting
$$\lambda_c=\kappa_c-{a_1\over a_2}\ ,\qquad\qquad \Delta=\lambda_c-\lambda\ ,
$$
we have
$${\partial \over \partial\Delta}(a_1-c)^{3/2}={a_2^{5/2}\over a_3}
\ .$$
Therefore,
$$ a_1-c =\Delta^{2/3}{a_2^{5/3}\over a_3^{2/3}}
$$
Since $c= {\partial F_0\over \partial\lambda}$ we finally obtain
$$ {\partial^2 F_0 \over \partial\Delta^2}
=-{\partial c\over \partial\Delta}= {2\over 3}
{a_2^{5/3}\over a_3^{2/3}} \Delta^{-1/3}
$$
which indicates that $\gm=1/3$, in agreement with ref. [\Das].

Now let us proceed to the two-matrix model which describes an Ising spin
($p=3$) coupled to gravity,
$$ \eqalign{& Z=\int {\cal D} \Phi_1 {\cal D} \Phi_2 \, e^{-N \,
   \left[ S(\Phi_1)+ S(\Phi_2)+ k\Tr \Phi_1 \Phi_2
    \right]} \ ,    \cr
& S(\Phi) =
   \Tr \left( \Phi^2 -
     \lambda \Phi^4 \right) -
    {g \over 2 N} \left( \Tr \Phi^4 \right)^2\ . \cr }
\eqn\ising$$
If we set
$$
 c = \left\langle {1 \over N} \Tr \Phi_1^4 \right\rangle
 = \left\langle {1 \over N} \Tr \Phi_2^4 \right\rangle
\eqn\selfis$$
and make the substitution \Replace\ in eq. \ising, then we arrive
at an auxiliary two-matrix model with no touching interactions and
coupling constant $\kappa=\lambda+ gc$. From the fact that, for
a specially chosen $k$, such
a model has $\gamma=-1/3$ it follows that the self-consistency condition
has the form
$$ c=a_1-a_2 (\kappa_c-\kappa)+ a_3 (\kappa_c-\kappa)^{4/3}+\ldots
$$
where $a_i$ and $\kappa_c$
are a new set of positive constants.
If we fine tune $g=1/a_2$, then
$$ {\partial c\over \partial\lambda}= {3\over 4}
{a_2^2\over a_3 (\kappa_c-\kappa)^{1/3}}=
{3\over 4}
{a_2^{7/3}\over a_3 (a_1-c)^{1/3}}
\ .$$
Integrating this equation, we find
$$ a_1-c =\Delta^{3/4}{a_2^{7/4}\over a_3^{3/4}}
\ .$$
Since ${\partial F_0\over \partial\lambda}=2c$, we finally obtain
$$ {\partial^2 F_0 \over \partial\Delta^2}=
{3\over 2}
\Delta^{-1/4}{a_2^{7/4}\over a_3^{3/4}}
$$
which shows that $\gm=1/4$, in agreement with eq. \ft.

It is now clear how to generalize our methods to an arbitrary two-matrix
model describing the $(p, p+1)$ minimal model coupled to gravity [\kkd],
$$ \eqalign{& Z=\int {\cal D} \Phi_1 {\cal D} \Phi_2 \, e^{-N \,
   \left[ S_p(\Phi_1)+ S_p(\Phi_2)+ k\Tr \Phi_1 \Phi_2
    \right]} \ ,    \cr
& S_p(\Phi) =
   \Tr \left( \Phi^2 -
     \lambda \Phi^4 +\ldots +\tau \Phi^{2p-2}\right) -
    {g \over 2 N} \left( \Tr \Phi^4 \right)^2\ . \cr }
\eqn\arbit $$
The parameters of the potential $S_p(\Phi)$ can be tuned
[\kkd] in such a way that,
for $g=0$,
$\gamma=-1/p$. For $g>0$, the effective quartic coupling is
$\kappa=\lambda+ gc$, where $c$ is defined by eq. \selfis.
The self-consistency condition has the form
$$ c=a_1-a_2 (\kappa_c-\kappa)+ a_3 (\kappa_c-\kappa)^{(p+1)/p}+\ldots
$$
If we fine-tune $g=1/a_2$ and repeat the familiar steps, we arrive at
$$ {\partial^2 F_0 \over \partial\Delta^2}\sim
\Delta^{-1/(p+1)}
$$
which implies $\gm = {1\over p+1}$, once again in agreement with eq. \ft.

The calculations in this section provide a check, in the context
of unitary minimal models coupled to gravity, of the
assertion that fine-tuning the touching interactions
makes the susceptibility exponent jump from $\gamma$ to
${\gamma\over \gamma-1}$. Ref. [\dur] and the \L\ gravity arguments of sec.
2 strongly suggest that this phenomenon is completely general.

\chapter{Discussion}

In this paper we have proposed a \L\ gravity formulation of the
\mm s with fine-tuned $ g(\Tr\Phi^n)^2$ terms.
It involves theories with $c\leq 1$ characterized by the unconventional
branch of gravitational dressing in the \L\ potential.
Thus, the hope that these theories correspond to $c>1$ is not realized.
Nevertheless, our proposal opens many directions for future research
which may shed new light on both \mm s and \L\ theory.
First, it would be interesting to see how our arguments square
with those of ref. [\ns].
Second, the calculation of string susceptibility
is only a first step in comparing the discretized and continuum approaches.
In fact, we should be able to extend most (if not all) the calculations
in conventional \mm s to the modified \mm s. The obvious questions are
correlation functions of scaling operators, higher-genus corrections,
etc. Although some results are available [\abc,\korch],
much progress
remains to be made. It is clearly of interest to carry out parallel
calculations in \L\ theory.

As remarked in sec. 2, some necessary \L\ gravity calculations are
extensions of known results, \eg\ the path integral at genus 1.
Furthermore, the spectrum of operators can be read off from this
path integral [\BKN].
Every operator can be written as ${\cal O} e^{\alpha\phi}$
where ${\cal O}$ depends on the matter, ghosts and the non-zero modes of
$\phi$.
Using the techniques of ref. [\BKN], one finds that the
spectrum of dimensions
of ${\cal O}$ is the same as in the conventional
\L\ theory.
The only new feature is that the operator
appearing in the action receives the $\alpha_-$ dressing.
We believe that, without additional fine-tuning, all the other
operators receive $\alpha_+$ dressing.
However, we may attempt to change the branch of dressing
of other operators by further fine-tuning. It would be
of interest to look for such effects in the \mm s.

A good motivation for studying operators with $\alpha_-$ dressing
is that they arise in the black hole model [\Wit,\sw], where the interaction
term is essentially $\partial X \bar \partial X  e^{-\sqrt 2\phi}$.
It would be very interesting to formulate a \mm\ which describes
such a two-dimensional black hole. There has been a number
of proposals for such a \mm\ [\BHmm,\antal], but it seems that fine-tuning
the $ g(\Tr\Phi^n)^2$
terms offers some intriguing new possibilities.

\ack
I thank S. Dalley, M. Douglas and S. Shenker for discussions.
I am grateful to the Aspen Center for Physics
where most of this project was carried out.
This work was supported in part by DOE grant DE-FG02-91ER40671,
the NSF Presidential Young Investigator Award PHY-9157482,
James S. McDonnell Foundation grant No. 91-48,
and an A. P. Sloan Foundation Research Fellowship.

\refout
\bye